\documentclass[%
 reprint,
superscriptaddress,
nofootinbib,
nobibnotes,
 amsmath,
 amssymb,
 aps,
 prl,
]{revtex4-1}

\usepackage{subfigure}
\usepackage{graphicx}

\begin{document}

\title{Long-distance distribution of atom-photon entanglement at telecom wavelength}

\author{Tim van Leent}
\altaffiliation{These authors contributed equally}
\affiliation{Fakult{\"a}t f{\"u}r Physik, Ludwig-Maximilians-Universit{\"a}t M{\"u}nchen, Schellingstr. 4, 80799 M{\"u}nchen, Germany}
\affiliation{Munich Center for Quantum Science and Technology (MCQST), Schellingstr. 4, 80799 M{\"u}nchen, Germany}

\author{Matthias Bock}
\altaffiliation{These authors contributed equally}
\affiliation{Fachrichtung Physik, Universit{\"a}t des Saarlandes, Campus E2.6, 66123 Saarbr{\"u}cken, Germany}

\author{Robert Garthoff}
\altaffiliation{These authors contributed equally}
\affiliation{Fakult{\"a}t f{\"u}r Physik, Ludwig-Maximilians-Universit{\"a}t M{\"u}nchen, Schellingstr. 4, 80799 M{\"u}nchen, Germany}
\affiliation{Munich Center for Quantum Science and Technology (MCQST), Schellingstr. 4, 80799 M{\"u}nchen, Germany}

\author{Kai Redeker}
\affiliation{Fakult{\"a}t f{\"u}r Physik, Ludwig-Maximilians-Universit{\"a}t M{\"u}nchen, Schellingstr. 4, 80799 M{\"u}nchen, Germany}
\affiliation{Munich Center for Quantum Science and Technology (MCQST), Schellingstr. 4, 80799 M{\"u}nchen, Germany}

\author{Wei Zhang}
\affiliation{Fakult{\"a}t f{\"u}r Physik, Ludwig-Maximilians-Universit{\"a}t M{\"u}nchen, Schellingstr. 4, 80799 M{\"u}nchen, Germany}
\affiliation{Munich Center for Quantum Science and Technology (MCQST), Schellingstr. 4, 80799 M{\"u}nchen, Germany}

\author{Tobias Bauer}
\affiliation{Fachrichtung Physik, Universit{\"a}t des Saarlandes, Campus E2.6, 66123 Saarbr{\"u}cken, Germany}

\author{Wenjamin Rosenfeld}
\affiliation{Fakult{\"a}t f{\"u}r Physik, Ludwig-Maximilians-Universit{\"a}t M{\"u}nchen, Schellingstr. 4, 80799 M{\"u}nchen, Germany}
\affiliation{Munich Center for Quantum Science and Technology (MCQST), Schellingstr. 4, 80799 M{\"u}nchen, Germany}
\affiliation{Max-Planck Institut f{\"u}r Quantenoptik, Hans-Kopfermann-Str. 1, 85748 Garching, Germany}

\author{Christoph Becher}
\altaffiliation{christoph.becher@physik.uni-saarland.de}
\affiliation{Fachrichtung Physik, Universit{\"a}t des Saarlandes, Campus E2.6, 66123 Saarbr{\"u}cken, Germany}
\email{christoph.becher@physik.uni-saarland.de}

\author{Harald Weinfurter}
\altaffiliation{h.w@lmu.de}
\affiliation{Fakult{\"a}t f{\"u}r Physik, Ludwig-Maximilians-Universit{\"a}t M{\"u}nchen, Schellingstr. 4, 80799 M{\"u}nchen, Germany}
\affiliation{Munich Center for Quantum Science and Technology (MCQST), Schellingstr. 4, 80799 M{\"u}nchen, Germany}
\affiliation{Max-Planck Institut f{\"u}r Quantenoptik, Hans-Kopfermann-Str. 1, 85748 Garching, Germany}
\email{h.w@lmu.de}

\date{\today}

\begin{abstract}
Entanglement between stationary quantum memories and photonic channels is the essential resource for future quantum networks. Together with entanglement distillation it will enable for efficient distribution of quantum states. Here we report on the generation and observation of entanglement between a Rb-87 atom and a photon at telecom wavelength over 20 km optical fiber. For this purpose, we use polarization-preserving quantum frequency conversion to transform the wavelength of a photon entangled with the atomic spin state from 780 nm to the telecom S-band at 1522 nm. We achieve an unprecedented external device conversion efficiency of 57\% and observe an entanglement fidelity between the atom and telecom photon of $\geqslant$78.5$\pm$0.9\% over 20 km optical fiber, mainly limited by decoherence of the atomic state. This result is an important milestone on the road to distribute quantum information on a large scale.
\end{abstract}

\maketitle

\textit{Introduction.---}Quantum repeaters will allow for scalable quantum networks \cite{briegel1998quantum, kimble2008quantum}, which are essential for large scale quantum communication and distributed quantum computing. In such networks, photon mediated entanglement is distributed among quantum memories at stationary nodes. Various candidates exist to serve as quantum memory, which also provide the light-matter interface in these nodes, for example trapped neutral atoms \cite{volz2006observation, ritter2012elementary}, atomic ensembles \cite{lukin2000entanglement,kuzmich2003generation,sangouard2011quantum}, trapped ions \cite{blinov2004observation, duan2010colloquium}, NV centers \cite{kurtsiefer2000stable,kalb2017entanglement}, quantum dots \cite{santori2002indistinguishable,childress2005fault}, or rare earth ions in solids \cite{kutluer2017solid,laplane2017multimode}.

For single atoms, several critical capabilities required to serve as a network node have recently been demonstrated. This concerns in particular, atom-atom quantum logic gates \cite{welte2018photon,saffman2010quantum,PhysRevLett.104.010502}, long qubit storage times \cite{korber2018decoherence}, high-fidelity heralded entanglement over hundreds of meters \cite{hofmann2012heralded,rosenfeld2017event}, and scalability of the number of individually addressable trapped atoms \cite{endres2016atom,barredo2016atom,de2019defect}. These achievements make single trapped atoms a promising candidate to develop a first quantum repeater link.

A capability not demonstrated for single atoms so far, however, is the distribution of entanglement at telecom wavelengths, which is indispensible for long, fiber-based quantum network links. By employing quantum frequency conversion (QFC) to telecom wavelength \cite{zaske2012visible, ates2012two, de2012quantum, ikuta2011wide}, the attenuation in fibers can be minimized while enabling the use of existing telecommunication infrastructure to economically realize network links. Recently, such concepts have been used to demonstrate entanglement between a telecom photon and an atomic ensemble \cite{dudin2010entanglement,ikuta2018polarization,yu2019entanglement}, a trapped ion \cite{bock2018high,Krutyanskiy2019lightNew}, or an NV center \cite{PhysRevLett.123.063601}.

Here we report on generation and detection of entanglement between a Rb-87 atom and a photon at telecom wavelength over long fiber links. The scheme starts with entangling the atomic spin state with the polarization state of a spontaneously emitted photon \cite{volz2006observation}. Subsequently, the wavelength of the photon is converted to the telecom S-band while preserving its polarization state and transferred over several km of optical fiber \cite{bock2018high}. The entanglement is analyzed by measuring atom-photon state correlations in two bases. Based on our results, we analyze the applicability of this scheme for long distance quantum links.

\begin{figure*}
\includegraphics[width=0.95\linewidth]{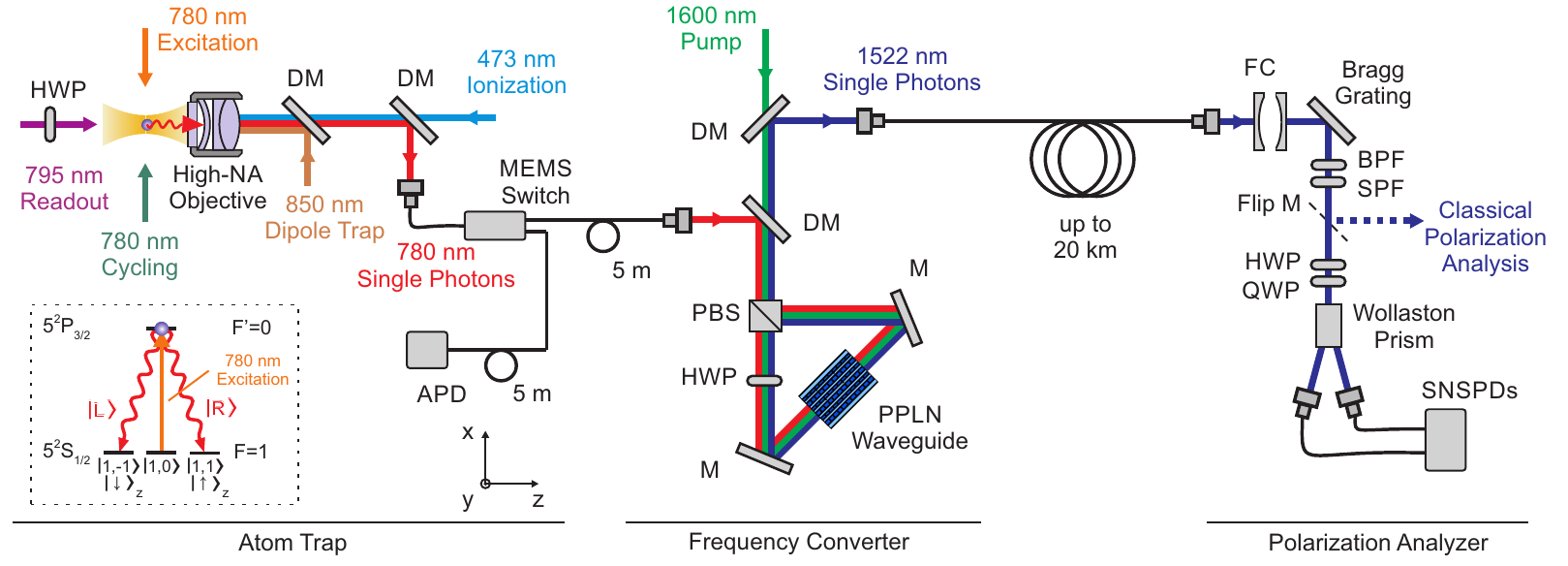}
\caption{\protect\textbf{Experimental setup and entanglement generation scheme.} A single Rb-87 atom (upper left side), serving as quantum memory, is stored in the focus of a dipole trap (850 nm wavelength, 2.05 $\mu$m waist, and 42 mW power), where a high-NA objective collects the atomic fluorescence. The atom-photon entanglement is generated in the spontaneous decay following the excitation to the state $5^2 P_{3/2}$ $|F'=0, m_{F'} = 0 \rangle$. The MEMS switch guides the emitted photons towards the frequency converter where the 780 nm single photons are overlapped with 1600 nm pump light within a PPLN waveguide in a Sagnac-type interferometer to transfer the entanglement to 1522 nm photons. In the polarization analyzer the single photons are first spectrally filtered by a Fabry-Perot filter cavity (FC), volume Bragg grating, bandpass filter (BPF), and shortpass filter (SPF). Next, after setting the analysis basis with a half-wave plate (HWP) and a quater-wave plate (QWP) and splitting the polarization components by a Wollaston prism, the single photons are detected with two SNSPDs. Classical reference light, inserted along the readout path, can be coupled out with a flip mirror to analyze and compensate for polarization drifts. Further abbreviations used: mirror (M) and dichroic mirror (DM).}
\label{fig:setup}
\end{figure*}

\textit{Methods.---}The experimental setup consists of a single atom trap with high-NA optics to collect the atomic fluorescence, a polarization-preserving quantum frequency converter in Sagnac configuration, and a polarization analyzer which is connected via single mode fiber of different lengths. For details see Fig. \ref{fig:setup}.

The experimental sequence starts by storing a single Rb-87 atom in an optical dipole trap \cite{volz2006observation,hofmann2012heralded} and preparing it in the initial state $5^2 S_{1/2}$ $|F=1, m_F = 0 \rangle$ via optical pumping. Next, a short laser pulse (22 ns FWHM) excites the atom to the state $5^2 P_{3/2}$ $|F'=0, m_{F'} = 0 \rangle$ (Fig. \ref{fig:setup}). In the subsequent spontaneous decay, the atomic spin state becomes entangled with the polarization state of the photon emitted along the quantization (z-) axis. This results in the following maximally entangled atom-photon state

\begin{equation}
\begin{split}
|\Psi \rangle_{\text{Atom-Photon}} & = \frac{1}{\sqrt{2}} (|\downarrow \rangle_z |L \rangle + |\uparrow \rangle_z |R \rangle) \\
 & = \frac{1}{\sqrt{2}} (|\downarrow \rangle_x |V \rangle + |\uparrow \rangle_x |H \rangle),
\end{split}
\end{equation}

\noindent where $|L\rangle$ and $|R\rangle$ denote left- and right-circular photonic polarization states, $|H\rangle$ and $|V\rangle$ denote the horizontal and vertical linear photonic polarization states, and $|\downarrow \rangle$ and $|\uparrow \rangle$ denote the atomic qubit state which, e.g. for the quantization axis z, corresponds to the states $|F=1, m_F = -1 \rangle$ or $|F=1, m_F = +1 \rangle$, respectively. 

A microelectromechanical systems (MEMS) switch is used to either guide the atomic fluorescence to a silicon avalanche photodiode (APD) during loading the trap or to the frequency converter during the state preparation and excitation cycles. These cycles are repeated until a single telecom photon is detected at the polarization analyzer, whereby the atom is cooled for 350 $\mu$s after each 40 excitations in order to minimize the thermal motion in the trap. The excitation rate is mainly limited by the travel time of the photon through the optical fiber, resulting in an average excitation rate of 7.3 kHz for the 20 km fiber.

To obtain the necessary performance for the presented experiment, the coherence time of the atomic state is prolonged to hundreds of $\mu$s by suppressing small ($\sim$mG) magnetic field fluctuations with a 42 mG constant magnetic field along the y-axis. Currently, the dominant decoherence effect is position-dependent dephasing caused by longitudinal field components of the strongly focused dipole trap \cite{coherence2019, burchardt2017thesis}. On the efficiency side, by implementing a custom designed high-NA objective, an overall efficiency of about $7.5\times10^{-3}$ is achieved without conversion when detecting with the APD ($\sim$55\% efficiency) after an excitation attempt. 

\begin{figure}
\centering
\includegraphics[width=0.6\linewidth]{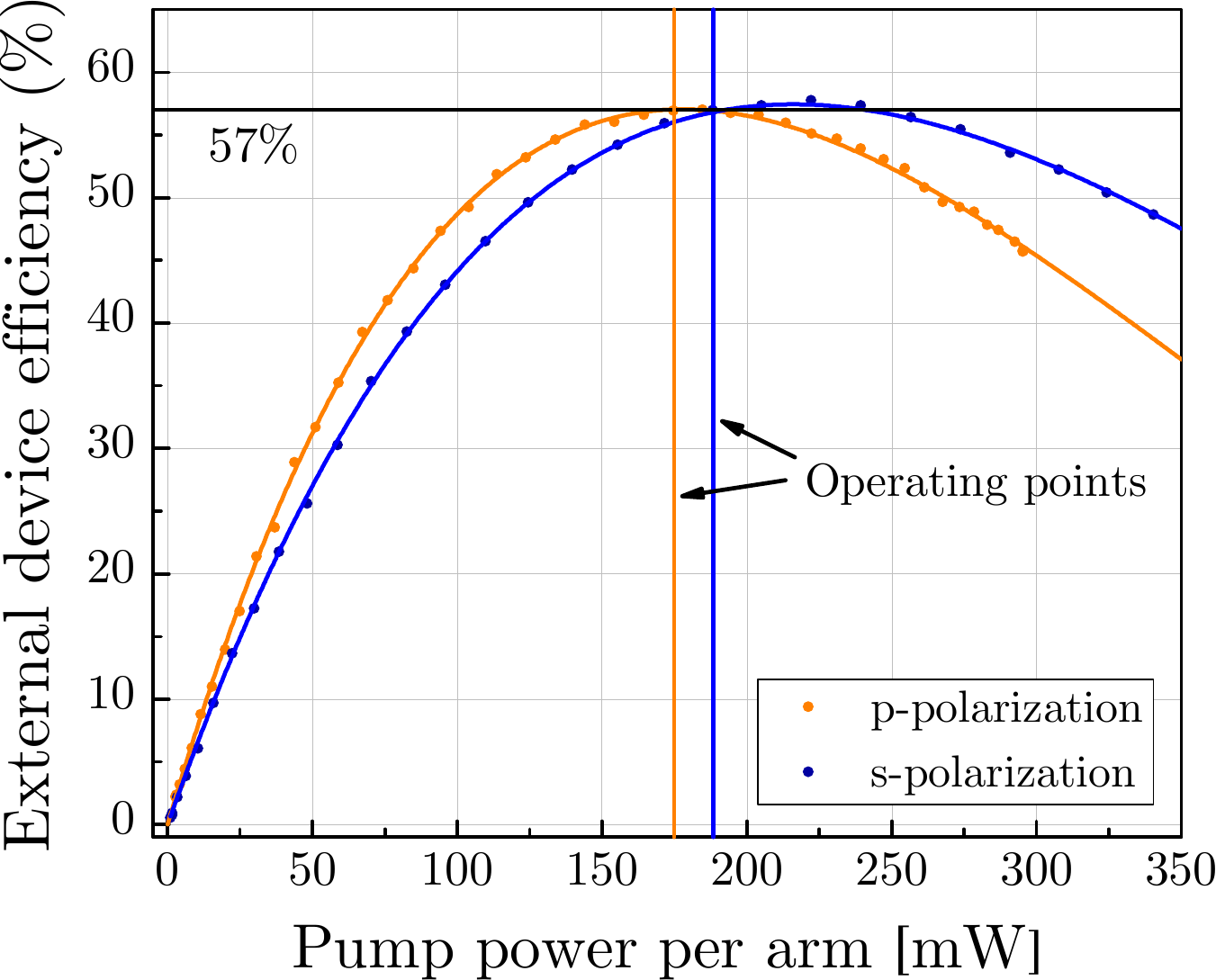}
\caption{\textbf{External device efficiency of the frequency converter.} The external device efficiency ($\eta$) of the two polarization components depends on the pump power ($P$) in the respective arm. The data are fitted with $\eta(P) = \eta_{\text{max}} \sin^2(\sqrt{\eta_{\text{nor}}P}L)$ \cite{zaske2011efficient}. The power in each arm is set to the operating point such that both conversion efficiencies are equal and one efficiency is maximized, 175 mW and 189 mW for the p- and s-polarization arm, respectively. At this point, the external device efficiency equals 57\%.}
\label{fig:conversion}
\end{figure}

The polarization-preserving QFC to the telecom S-band at 1522 nm is realized by difference-frequency generation (DFG) in a periodically poled lithium niobate (PPLN) waveguide. To this end, the single photons at 780 nm are mixed with a strong cw pump field at 1600 nm within a waveguide in a Sagnac-type setup \cite{ikuta2018polarization}. The input polarization is split into two arms with a polarizing beam-splitter (PBS), whereby a half-wave plate (HWP) is introduced in one of the arms such that the two counterpropagating beams have the same polarization when entering the waveguide. The conversion efficiency of both arms is set equal by fine tuning the pump field powers, see Fig. \ref{fig:conversion}.

\begin{figure*}
	\centering
	\subfigure[]{\includegraphics[width=.3\textwidth]{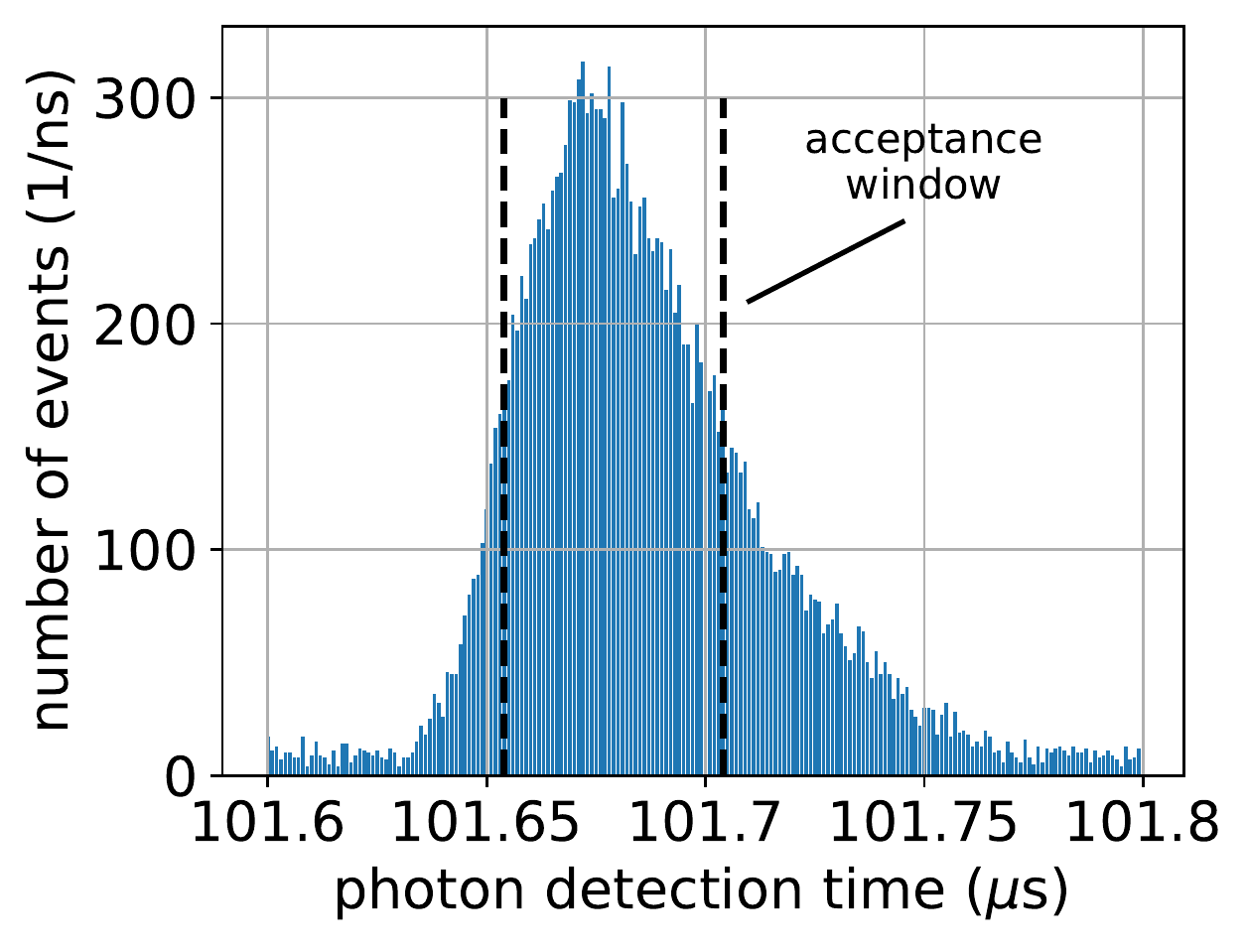}}\quad 
	\subfigure[]{\includegraphics[width=.6\textwidth]{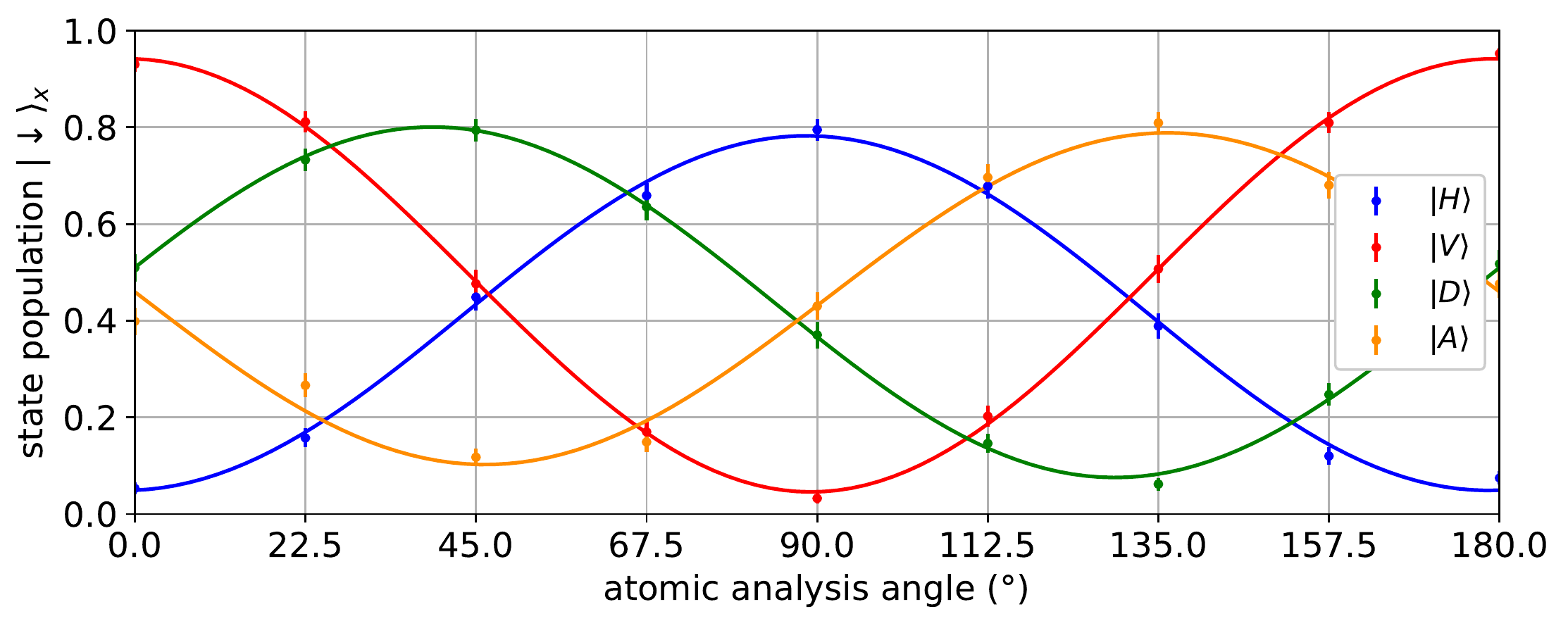}}
	\caption{\textbf{Observation of atom-photon entanglement over 20 km optical fiber ($A$).} (a) Detection time histogram of the frequency converted photons. Within a hardwired acceptance window of 50 ns, indicated with dashed lines, approximately 2/3 of the converted single photons were accepted. Note that the QFC does not influence the photon shape \cite{zaske2012visible}, see \cite{hofmann2012heralded} for an unconverted photon shape. (b) The corresponding atom-photon state correlations in two bases (H/V and D/A) for varying atomic analysis angle (i.e. readout polarization, whereby 0$^{\circ}$ corresponds to vertical polarization). The sinusoidal fits give estimated visibilities of 73.4$\pm$2.0\%, 89.6$\pm$1.1\%, 72.5$\pm$1.1\%, and 68.6$\pm$4.1\% for horizontal $|H\rangle$, vertical $|V\rangle$, diagonal $|D\rangle$, and anti-diagonal $|A\rangle$ photonic linear polarization states, respectively. This results in an estimated state fidelity of $\geqslant$78.5$\pm$0.9\%.}
	\label{fig:results}
\end{figure*}

Various spectral filtering stages efficiently separate the single telecom photon from the strong pump field and the noise induced by the pump field via anti-stokes Raman scattering in the waveguide. For this purpose, we introduce a combination of two shortpass filters (SPF) with a cut-off at 1560 nm, a bandpass filter (BPF) at 1535 nm with a bandwith of 30 nm FWHM, a volume Bragg grating (25 GHz FWHM), and a Fabry-Perot filter cavity (FC) with a finesse of 700 and a bandwidth of 27 MHz FWHM. The latter is locked to the pump laser at 1600 nm while having a further resonance at the telecom single photon frequency. See Supplemental Material \cite{SM} Sec. I for more details about the QFC.

The external device efficiency of the frequency converter, defined from the input fiber of the converter until the first waveplate of the polarization analyzer and including a 50 m optical fiber, equals 57\%. This is, to the best of our knowledge, the highest external device efficiency reported so far. The efficiency is limited by the transmission through optical elements (82.6\%), fiber coupling (87.8\%), waveguide coupling (90.0\%), and spectral filtering (90.7\%), which reduce the high internal conversion efficiency of 96.2\%. 

The polarization state of the telecom photons is analyzed after propagating through up to 20 km optical fiber (SMF-28). Single photons are detected by two superconducting nanowire single-photon detectors (SNSPDs) with efficiencies of 16\% and 18\% at 1522 nm. Detection events are accepted within a hardwired interval of 50 ns, which results in accepting approximately 2/3 of the converted single photons. Polarization rotations in the complete single photon beam path are analyzed and can be compensated for by using classical reference light and a fiber polarization controller \cite{rosenfeld2017event}. 

Following a successful photon detection event, the atomic spin state is analyzed using a state-selective ionization scheme. The state selectivity is controlled by the polarization of the readout pulse, see Supplemental Material \cite{SM} Sec. II for more details.

\textit{Results.---}We analyzed the entanglement between the atom and photon at telecom wavelength after a 20 km ($A$), 10 km ($B$), and 50 m ($C$) optical fiber. Furthermore, to investigate possible noise induced by the QFC, a reference measurement was performed without frequency conversion ($D$). For measurements $C$ and $D$ we applied an atomic state detection delay of 50 $\mu$s to make the loss in fidelity due to atomic state decoherence comparable to measurement $B$. All results are summarized in Tab. \ref{tab:results}. 

\begin{table}
\caption{\textbf{Observation of atom-photon entanglement for different experimental configurations.} The measurement configurations $A$, $B$, $C$, and $D$ differ in optical fiber length, detection wavelength, and/or atomic state detection delay. The fidelity is obtained via Eq. \ref{eq:fidelity}, whereby the visibilities are fitted from the measured correlation probabilities. The $S$ parameter (CHSH) is calculated directly from the measured Bell states.}
\bgroup
\def\arraystretch{1.3}
\begin{tabular}{ l | c | c | c | c  }
 & ($A$)  & ($B$) & ($C$) & ($D$) \\
fiber length & 20 km & 10 km & 50 m & 5 m \\
wavelength & 1522 nm &  1522 nm & 1522 nm  & 780 nm \\
readout delay &  102 $\mu$s &  51 $\mu$s & 51 $\mu$s & 51 $\mu$s \\ \hline 
fidelity (\%) & 78.5$\pm$0.9 & 84.3$\pm$0.9 & 88.0$\pm$0.8 & 89.7$\pm$0.7 \\
$S$ (CHSH) & 2.12$\pm$0.05 & 2.37$\pm$0.04 & 2.41$\pm$0.03 & 2.49$\pm$0.03    \\
SNR & 25.1 & 23.2 & 32.3 & 934.2  
\end{tabular}
\egroup
\label{tab:results}
\end{table}

A photon detection time histogram and corresponding atom-photon state correlations over 20 km optical fiber are shown in Fig. \ref{fig:results}a and b, respectively. In this measurement, 11335 events were observed within 360 minutes with an overall efficiency of detecting a telecom photon after an atomic excitation pulse of $0.173 \times 10^{-3}$. When optimizing the experiment for efficiency, e.g. by employing efficient single photon detectors ($>$85\%) and replacing the lossy MEMS switch (25\% loss), we expect an improvement of the overall efficiency by about one order of magnitude. The event rate in all measurements is $\sim$35 per minute, mainly limited by the atom loading time of about 1 s since the atom is lost during the state readout process in approximately half of the cases.

The signal-to-noise ratio (SNR) of the photon detection is mainly limited by the QFC pump laser induced noise and detector dark-counts, see Supplemental Material \cite{SM} Sec. I for details. For measurement $A$ these contributions amounted to 128 and 18 cps summed over both detectors, respectively. The SNR of 25.1 corresponds to our expectation, taking into account the 50 ns photon acceptance interval and the overall efficiency mentioned above. Variations in the SNR between measurements $A$, $B$, and $C$ originate from different fiber lengths as well as from slight laser power fluctuations in the atomic state preparation and excitation cycles.

To analyze the entanglement, the photonic polarization state was measured in the H/V (horizontal/vertical) and D/A (diagonal/anti-diagonal) basis, while varying the atomic analysis angle (i.e. readout polarization). The visibilities of the measured states are obtained by fitting the data with sinusoidal curves. The average visibility ($\bar{V}$) of the entangled state is estimated by assuming that the visibility in the third (unmeasured) basis is equal to the D/A basis. This results in estimated average visibilities of 74.2$\pm$1.0\%, 81.2$\pm$1.1\%, 85.6$\pm$0.9\%, and 87.4$\pm$0.6\% for measurement $A$, $B$, $C$, and $D$, respectively. Note that the visibility of the detected photon state $|V\rangle$ (e.g. red curve in Fig. \ref{fig:results}b) is significantly higher than the other states since the resulting atomic state is insensitive to the position-dependent dephasing of the atomic state, which is the dominant decoherence effect for all other states.

To compute a fidelity based on the measured visibilities one needs to consider that the atom is a spin-1 system. Hence, also a third atomic spin state can be populated ($5^2 S_{1/2}$ $|F=1, m_F = 0 \rangle$). Imperfections in the experiment, such as small magnetic fields ($\sim$mG) in direction not coinciding with the quantization axis, can lead to a population in this state. Accordingly, assuming isotropic dephasing towards white noise in the 2x3 state space, a lower bound on the fidelity of the entangled state is given by

\begin{equation}
F \geqslant \frac{1}{6} + \frac{5}{6} \bar{V},
\label{eq:fidelity}
\end{equation}

\noindent which results in fidelities of $\geqslant$78.5$\pm$0.9\%, 84.3$\pm$0.9\%, 88.0$\pm$0.8\%, and 89.7$\pm$0.7\% relative to a maximally entangled state for measurement $A$, $B$, $C$, and $D$, respectively. For measurement $A$, contributions to the loss in fidelity are the imperfect atomic state readout (3\%), atomic state decoherence (11\%), SNR in the photon detection (4\%), and experimental drifts (3\%).

The influence of the QFC on the state fidelity is best analyzed by comparing measurement $C$ (88.0$\pm$0.8\%) and $D$ (89.5$\pm$0.5\%), since the experimental configurations are most similar. For these measurements, the difference in fidelity can be solely explained by their difference in SNR (3\%).

When comparing measurement $B$ (84.3$\pm$0.9\%) and $C$ (88.0$\pm$0.8\%), the difference in fidelity is more than what can be expected from the lower SNR alone (1\%). We attribute the additional fidelity loss (3\%) in measurement $B$ to instabilities in the experiment, such as fiber polarization, magnetic field, and laser power drifts. The difference in fidelity between measurement $A$ (78.5$\pm$0.9\%) and $B$ (84.3$\pm$0.9\%) originates primarily from atomic state decoherence (5\%). 

All four measurements also include two setting combinations for a CHSH Bell test \cite{chsh1969}. The CHSH inequality requires two measurement settings for the photonic state (H/V \& D/A) and two measurement settings for the atomic state (22.5$^{\circ}$ or 157.5$^{\circ}$ \& 67.5$^{\circ}$ or 112.5$^{\circ}$). The resulting $S$ parameters, all clearly violating the CHSH-Bell inequality, are listed in Tab. \ref{tab:results}.

\textit{Outlook.---}The next milestone towards large scale quantum networks is to distribute heralded entanglement at telecom wavelength between two distant atomic memories via the entanglement swapping protocol \cite{hofmann2012heralded}. For this purpose, it is necessary to introduce frequency conversion for a second atom trap and install a set of SNSPDs optimized for 1522 nm. 

The expected atom-atom entanglement fidelity for varying distance is estimated by $F_{a-a}\geqslant 1/9 + 8/9 \bar{V}$, where the average atom-atom visibility $\bar{V}$ is estimated by squaring the corresponding atom-photon visibility and taking into account a 94\% two-photon interference contrast \cite{hofmann2012heralded}. Furthermore, we assume that the measurement needed for the entanglement swapping is performed at a middle station such that the atomic coherence time needed to distribute atom-photon and atom-atom entanglement over the same distance evens up.

\begin{figure}
\centering
\includegraphics[width=0.9\linewidth]{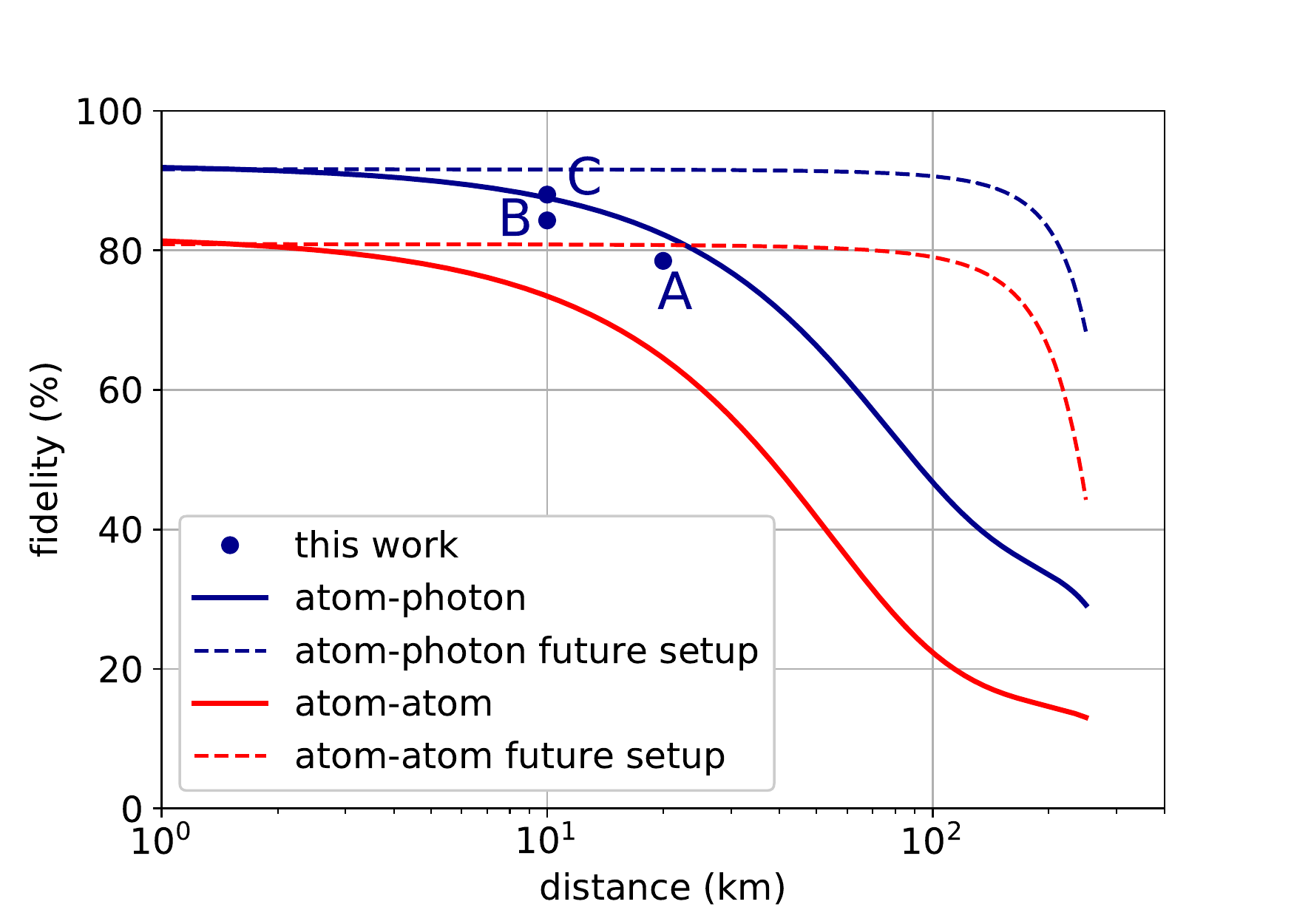}
\caption{\textbf{Expected atom-photon and atom-atom entanglement fidelities.} Points marked as $A$, $B$, and $C$ correspond to the measurements presented in the text and summarized in Tab. \ref{tab:results}. The expected fidelity for short distances ($<$1 km) is mainly limited by the imperfect atomic state preparation and readout. For distances $<$100 km, the fidelity is also reduced by atomic state decoherence due to position-dependent dephasing, which could be strongly suppressed in a future setup. For distances $>$100 km, detector dark-counts will eventually limit the fidelity.}
\label{fig:rates}
\end{figure}

Fig. \ref{fig:rates} shows the expected entanglement fidelities for a range of distances between the quantum memories. Below 1 km the performance is limited by imperfections in the atomic state preparation and analysis. For longer distances, the atomic state decoherence due to the position-dependent dephasing will significantly reduce the fidelity \cite{burchardt2017thesis}. A new trap geometry, involving a standing wave dipole trap, promises to strongly reduce this decoherence effect. The expected fidelities for this future setup are for large distances eventually limited by detector dark-counts. 

Using the current trap geometry, we expect an atom-atom entanglement fidelity of 65\% over 20 km with an event rate of 1 per minute. By improving the trap geometry, we expect this fidelity to increase to 81\%. Together with entanglement distillation \cite{kalb2017entanglement}, this allows for efficient distribution of quantum states over long distances.

\textit{Conclusion.---}Thanks to the record high external device efficiency of 57\% for the polarization-preserving QFC and improved collection optics for the atomic fluorescence, atom-photon entanglement was distributed and observed at high rate with a fidelity of at least 78.5$\pm$0.9\% over 20 km optical fiber. Implementing realistic improvements and extrapolating to even longer distances shows that entanglement distribution between atomic memories is feasible with a fidelity of more than 80\% over up to 100 km, thereby forming a valuable component for future quantum repeater based quantum networks. \\

\begin{acknowledgments}
We thank Benjamin Kambs and Stephan Kucera for valuable discussions regarding the QFC, Harald Giessen for lending us the 10 km fiber, and Matthias Kreis for technical support on the cavity locking electronics. We acknowledge funding by the German Federal Ministry of Education and Research (Bundesministerium f{\"u}r Bildung und Forschung (BMBF)) within the projects Q.com.Q and Q.Link.X (Contracts No. 16KIS0127, 16KIS0123, 16KIS0864, and 16KIS0880) and by the Deutsche Forschungsgemeinschaft (DFG, German Research Foundation) under Germany’s Excellence Strategy – EXC-2111 – 390814868.
\end{acknowledgments}

\bibliographystyle{apsrev4-1}

\begin{thebibliography}{39}
\bibitem [{coh()}]{coherence2019}%
  \BibitemOpen
  \href@noop {} {\bibinfo  {journal} {To be published}\ }\BibitemShut {NoStop}%
\end{thebibliography}


%

\end{document}


\title{Supplemental Material for:\\ Long-distance distribution of atom-photon entanglement at telecom wavelength}

\author{Tim van Leent}
\altaffiliation{These authors contributed equally}
\affiliation{Fakult{\"a}t f{\"u}r Physik, Ludwig-Maximilians-Universit{\"a}t M{\"u}nchen, Schellingstr. 4, 80799 M{\"u}nchen, Germany}
\affiliation{Munich Center for Quantum Science and Technology (MCQST), Schellingstr. 4, 80799 M{\"u}nchen, Germany}

\author{Matthias Bock}
\altaffiliation{These authors contributed equally}
\affiliation{Fachrichtung Physik, Universit{\"a}t des Saarlandes, Campus E2.6, 66123 Saarbr{\"u}cken, Germany}

\author{Robert Garthoff}
\altaffiliation{These authors contributed equally}
\affiliation{Fakult{\"a}t f{\"u}r Physik, Ludwig-Maximilians-Universit{\"a}t M{\"u}nchen, Schellingstr. 4, 80799 M{\"u}nchen, Germany}
\affiliation{Munich Center for Quantum Science and Technology (MCQST), Schellingstr. 4, 80799 M{\"u}nchen, Germany}

\author{Kai Redeker}
\affiliation{Fakult{\"a}t f{\"u}r Physik, Ludwig-Maximilians-Universit{\"a}t M{\"u}nchen, Schellingstr. 4, 80799 M{\"u}nchen, Germany}
\affiliation{Munich Center for Quantum Science and Technology (MCQST), Schellingstr. 4, 80799 M{\"u}nchen, Germany}

\author{Wei Zhang}
\affiliation{Fakult{\"a}t f{\"u}r Physik, Ludwig-Maximilians-Universit{\"a}t M{\"u}nchen, Schellingstr. 4, 80799 M{\"u}nchen, Germany}
\affiliation{Munich Center for Quantum Science and Technology (MCQST), Schellingstr. 4, 80799 M{\"u}nchen, Germany}

\author{Tobias Bauer}
\affiliation{Fachrichtung Physik, Universit{\"a}t des Saarlandes, Campus E2.6, 66123 Saarbr{\"u}cken, Germany}

\author{Wenjamin Rosenfeld}
\affiliation{Fakult{\"a}t f{\"u}r Physik, Ludwig-Maximilians-Universit{\"a}t M{\"u}nchen, Schellingstr. 4, 80799 M{\"u}nchen, Germany}
\affiliation{Munich Center for Quantum Science and Technology (MCQST), Schellingstr. 4, 80799 M{\"u}nchen, Germany}
\affiliation{Max-Planck Institut f{\"u}r Quantenoptik, Hans-Kopfermann-Str. 1, 85748 Garching, Germany}

\author{Christoph Becher}
\altaffiliation{christoph.becher@physik.uni-saarland.de}
\affiliation{Fachrichtung Physik, Universit{\"a}t des Saarlandes, Campus E2.6, 66123 Saarbr{\"u}cken, Germany}
\email{christoph.becher@physik.uni-saarland.de}

\author{Harald Weinfurter}
\altaffiliation{h.w@lmu.de}
\affiliation{Fakult{\"a}t f{\"u}r Physik, Ludwig-Maximilians-Universit{\"a}t M{\"u}nchen, Schellingstr. 4, 80799 M{\"u}nchen, Germany}
\affiliation{Munich Center for Quantum Science and Technology (MCQST), Schellingstr. 4, 80799 M{\"u}nchen, Germany}
\affiliation{Max-Planck Institut f{\"u}r Quantenoptik, Hans-Kopfermann-Str. 1, 85748 Garching, Germany}
\email{h.w@lmu.de}

\date{\today}

\maketitle

\onecolumngrid

\setcounter{equation}{0}
\setcounter{figure}{0}
\setcounter{table}{0}
\setcounter{section}{0}
\setcounter{page}{1}
\makeatletter
\renewcommand{\theequation}{S\arabic{equation}}
\renewcommand{\thefigure}{S\arabic{figure}}
\renewcommand{\bibnumfmt}[1]{[#1]}
\renewcommand{\citenumfont}[1]{#1}

\section{QFC system for a Rubidium-atom based quantum network link} 
\label{sec:QFC_sys}

The quantum frequency conversion (QFC) device utilized in the experiment described in the main text is part of a complete QFC system designed as telecom interface for extending the Munich elementary quantum network link \cite{RosenfeldPRL2017}. The conversion relies on the well-established $\chi^{(2)}$-nonlinear process of difference frequency generation (DFG) in which single photons at 780$\,$nm are combined with a cw-laser around 1600$\,$nm to achieve down-conversion to the telecom S-band at 1522$\,$nm. 

\subsection{Design criteria}

The particular wavelength combination has been selected to fulfill a series of criteria:

\begin{itemize}[leftmargin=1.2cm]
\item The target wavelength should be within the low-loss telecom bands between 1260$\,$nm and 1625$\,$nm; preferably as close as possible to the telecom C-band around 1550$\,$nm where attenuation in optical fibers is minimal.
\item The mixing laser wavelength for the DFG-process should be the longest wavelength to avoid noise generated by Stokes Raman scattering or non-phasematched spontaneous parametric down-conversion around the target wavelength ("long-wavelength pumping"). Moreover, the frequency difference between mixing and target wavelength should be maximized to reduce the anti-Stokes Raman noise. 
\item The frequency-converted photons should be indistinguishable with respect to their central wavelength as well as their spectral and temporal shape to allow for high-contrast quantum interference, a prerequisite to establish entanglement between remote quantum nodes via the entanglement swapping protocol. Preserving spectral and temporal properties of the input photons requires a single-frequency mixing laser with a linewidth substantially smaller than the photon bandwidth (6.1$\,$MHz). To obtain identical central wavelengths for various QFC systems, two options are apparent: using a single mixing laser which is distributed to the frequency converters located at each node or using separate mixing lasers for each converter whose frequencies are absolutely stabilized with a precision much smaller than the photon bandwidth.   
\item The mixing laser has to be spatially single-mode with an output power around 2$\,$W in order to achieve internal conversion efficiencies close to 100\%.
\item The mixing laser system has to be stable (with respect to e.g. its output power or its spectral properties), reliable, and fail-safe to enable 24/7-operation. This is considered necessary especially as complexity (and potentially also the measurement periods) increases along with an increasing number of nodes, hence requirements on the reliability of all components also increase.  
\end{itemize}  

Mixing lasers with wavelengths ranging from 1565$\,$nm to 2050$\,$nm can be used to convert an input wavelength of 780$\,$nm to the telecom bands when restricted to the long-wavelength pumping regime. Furthermore, to allow for requirements on spectral properties, output power and long-term stability, a MOPA (master oscillator power amplifier) configuration comprised of a narrow-band, tunable master laser (e.g. single-frequency diode or fiber lasers) and a fiber amplifier is a fitting candidate for the mixing laser system. This cuts the wavelength range into two windows; one from 1565$\,$nm to 1605$\,$nm covered by Erbium-doped fiber amplifiers (EDFA) \cite{AlbrechtNatCom2014, IkutaNatcom2018} and one from 1900$\,$nm to 2050$\,$nm where Thulium-doped fiber lasers/amplifiers (TDFA) are available \cite{YuArxiv2019}. While the option with Thulium-based lasers offers superior noise properties in the QFC, Erbium-based systems are far more developed and allow target wavelengths in or close to the C-band. Thus we chose 1600$\,$nm as mixing wavelength, which is a tradeoff between the frequency separation between mixing and target wavelength (1522$\,$nm) and the width of the Erbium gain spectrum. Additionally, 1600$\,$nm can be transmitted with low losses through optical fibers, i.e. even if the nodes are separated by several kilometers, a single master laser in combination with EDFAs is enough to supply all QFC devices. Hence spectral indistinguishability of the frequency-converted photons is guaranteed.

The QFC system is designed for comparatively easy relocation as well as easy integration into existing quantum optics experiments. To this end, it consists of a series of mobile standalone platforms, each featuring a top layer with the optical setups placed on honeycomb breadboards, and a housing underneath with at least one layer containing control electronics. To reduce mechanical vibrations, all platforms are damped with sorbothane isolators.

The system  consists in total of four platforms, in particular, a master laser system to provide frequency-stabilized light for the DFG-process, two polarization-preserving QFC devices to equip both nodes in the network link with a telecom interface, and a platform to perform a Bell state measurement (BSM) at telecom wavelengths.

The experiment in the main text necessitated only the master laser system, one QFC device, and a polarization analyzer, which is realized by one branch of the BSM setup. Those platforms are presented in the following sections in more detail.   

\subsection{Master laser system}  
\label{sec:MaLa}

\begin{figure}[h]
\centering
\includegraphics[width=0.65\linewidth]{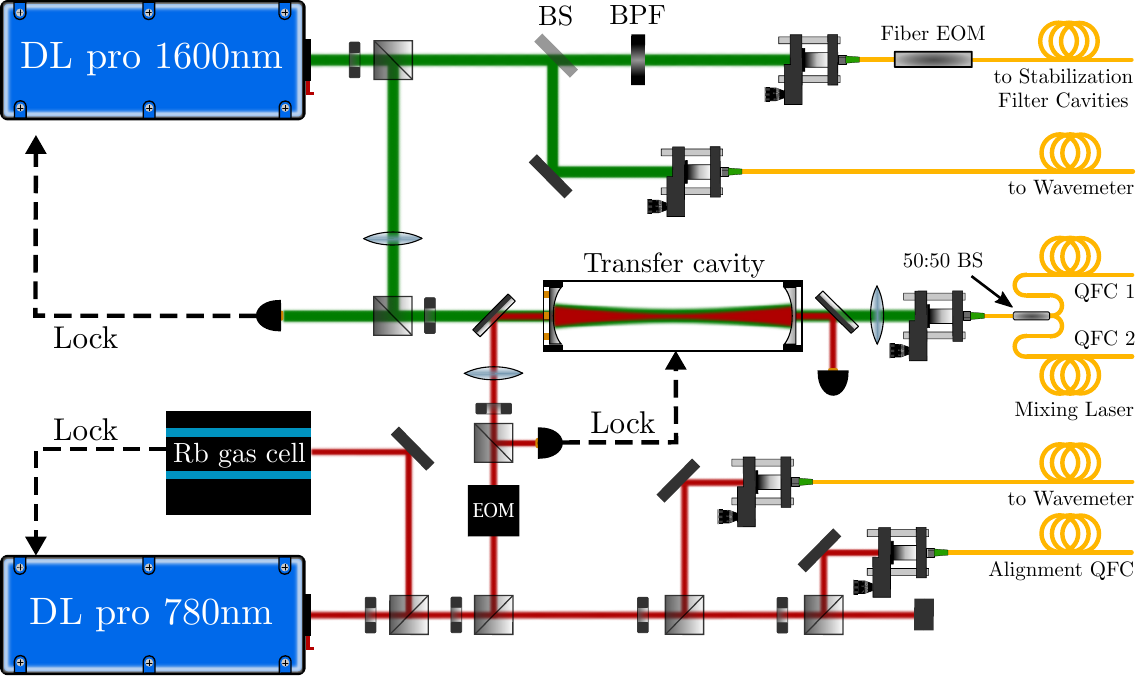}
\caption{\textbf{Setup of the master laser system.} Details on the setup can be found in the text. The following abbreviations are used: beam-splitter (BS), bandpass filter (BPF), and electro-optic modulator (EOM).}
\label{fig:lasersystem}
\end{figure}

The operation of the QFC system requires two lasers; one at the input wavelength (780$\,$nm) and one at the mixing laser wavelength (1600$\,$nm). As described in the main text, narrow-band spectral filtering of the frequency-converted photons down to 27$\,$MHz is applied to reduce noise at the single photon level; i.e. both lasers must be frequency-stabilized to an absolute long-term stability in the order of 1$\,$MHz. Hence, the master laser platform contains - in addition to the lasers - components for the frequency stabilization, control and stabilization electronics and optical components to distribute the lasers to the other platforms. The setup is shown in Fig. \ref{fig:lasersystem}.
  
Both lasers are tunable, single-frequency diode lasers (DL pro, \textit{Toptica Photonics}). The 780$\,$nm-laser is utilized to provide an absolute frequency reference to stabilize the 1600$\,$nm-laser and for alignment of the QFC devices. Its frequency is stabilized to the Rb-87 hyperfine transition $5^{2}S_{1/2}, F=1 \leftrightarrow 5^{2}P_{3/2}, F'=0$ (same transition as the emitted photons in the experiment) by doppler-free saturation spectroscopy of a gas cell (Cosy, \textit{TEM Messtechnik}).
        
The 1600$\,$nm-laser serves mainly as mixing laser in the DFG process. 
Due to the absence of narrow spectroscopy transitions ($<$ 1$\,$MHz), it is frequency stabilized at 1600.1421$\,$nm (corresponding to a difference frequency of 1522.7295$\,$nm) via a transfer cavity lock to the 780$\,$nm laser. To this end, we adapted the transfer cavity design from \cite{RohdeJPB2010}, wherby the confocal cavity has a free spectral range of 500$\,$MHz and a finesse of about 300. The cavity length is locked to the 780$\,$nm-laser by the Pound-Drever-Hall (PDH) technique using home-built electronics \cite{RohdeJPB2010}. Next, the mixing laser frequency is locked to the cavity---as well with a PDH lock---whereby the diode current is modulated to get sidebands at 12.5$\,$MHz. Since the sidebands are unwanted in the DFG-process and the diode laser occasionally gets slightly multimode, the transfer cavity is additionally utilized as narrow-band filter before the pump light is distributed to the QFC devices.
 
The rather small free spectral range (500$\,$MHz) of the transfer cavity necessitates an additional absolute reference at 1600$\,$nm. To this end, we utilize a wavelength meter (WS6-200, \textit{High Finesse}), which is calibrated with the 780$\,$nm-laser about 2-3 times per week to remove long-term drifts of the absolute accuracy due to environmental effects.

In addition we use the 1600$\,$nm-laser also for an active stabilization of the narrow-band filter cavities, which are designed to be doubly-resonant for the mixing laser and the converted light. Further details on this will be explained in Sec. \ref{sec:SpecFil}. 

\subsection{Polarization-preserving QFC device}

\begin{figure}[h]
\centering
\includegraphics[width=0.47\linewidth]{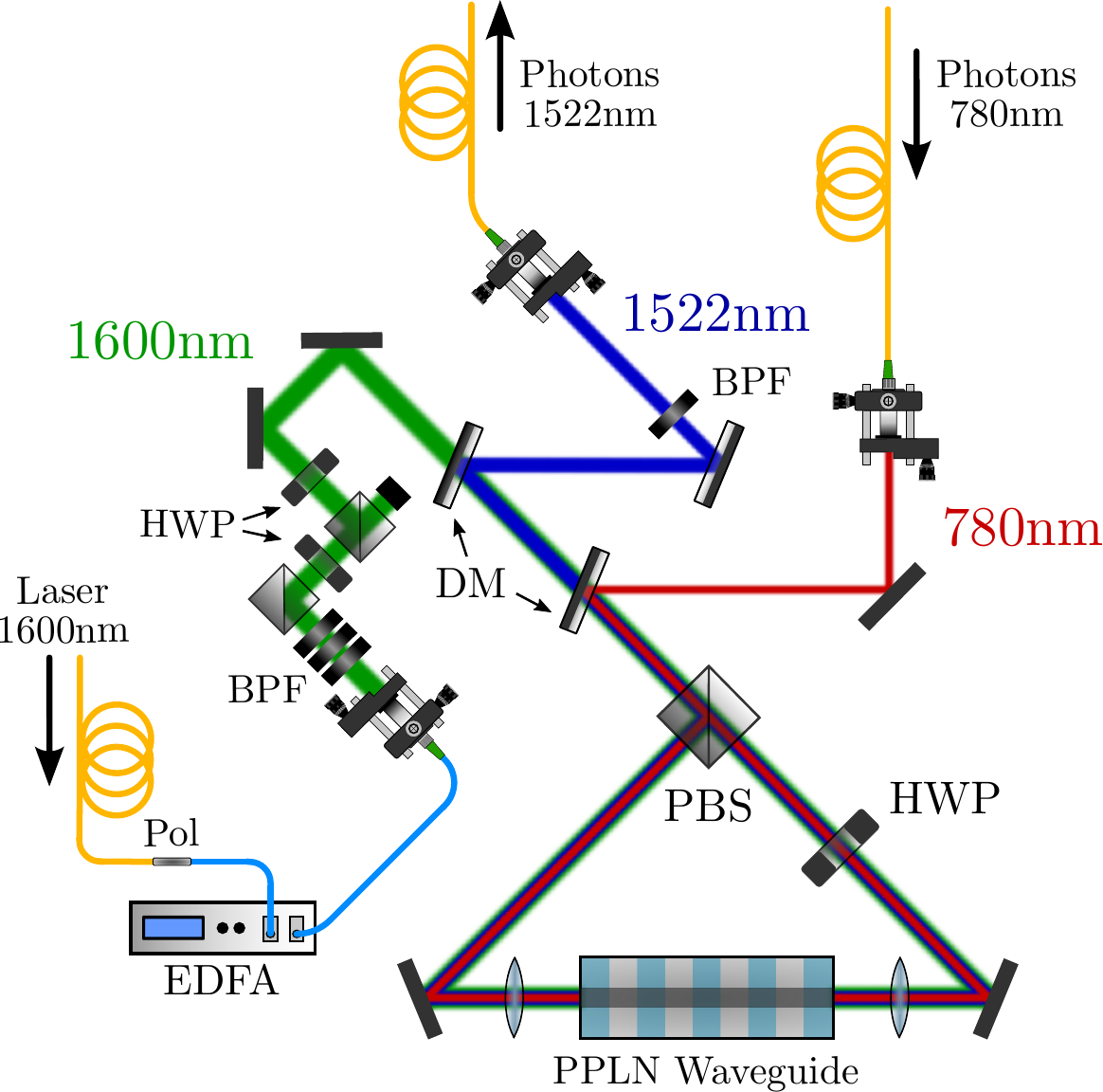}
\caption{\textbf{Setup of the QFC devices.} Details on the setup can be found in the text. The following abbreviations are used: polarizer (Pol), Erbium-doped fiber amplifier (EDFA), band-pass filter (BPF), half wave plate (HWP), dichroic mirror (DM), polarizing beam splitter (PBS), periodically poled lithium niobate (PPLN).}
\label{fig:qfc_det}
\end{figure}

The system's key element is a polarization-preserving QFC device; the experimental setup is shown in Fig. \ref{fig:qfc_det}. The conversion takes place in a temperature-stabilized periodically-poled lithium niobate (PPLN) ridge waveguide (\textit{NTT Electronics}) with 40$\,$mm length, lateral dimensions of around 9-10$\,\mu$m, and a poling period of $\Lambda$ = 18.25$\,\mu$m. To overcome the intrinsic polarization dependence of the DFG-process, several strategies were investigated and implemented over the last years \cite{KrutAPB2017, RamelowPRA2012, KaiserArxiv2019, BockNatCom2018, AlbotaJOS2006, IkutaNatcom2018}. Here we rely on a Sagnac-type configuration, which features - in contrast to Mach-Zehnder-type configurations - an intrinsic phase stability. Compared to our previous QFC device \cite{BockNatCom2018} this removes a lot of technical overhead related to the path length stabilization.

The input light at 780$\,$nm is coupled out of a single-mode fiber with an aspheric lens (f = 11$\,$mm) and overlapped with the pump laser at 1600$\,$nm on a dichroic mirror (DM). The latter is generated by an EDFA, which amplifies the weak mixing laser guided via a single-mode fiber from the master laser system to the QFC device. A fiber polarizer ensures that only correctly linear-polarized light is seeded into the EDFA. The amplified light (power around 1.2$\,$W) is coupled out of a polarization-maintaining fiber with an aspheric lens (f = 11$\,$mm) and spectrally filtered by a stack of three bandpass filters (center wavelength: 1600$\,$nm, FWHM: 50$\,$nm, \textit{Edmund optics}) to remove amplified spontaneous emission (ASE) noise at the target wavelength. Next, a polarizing beam splitter (PBS) cleans up the output light's polarization, a subsequent half-wave plate (HWP) and a PBS control the overall power and another HWP sets the relative power in the interferometer arms.

The Sagnac interferometer is formed by a PBS, a half wave plate (HWP), two silver mirrors, and the PPLN waveguide chip. PBS and HWP (both from \textit{B. Halle Nachfl.}) are achromatic, i.e. they work for all three wavelengths to ensure that all fields are s-polarized in the waveguide. Efficient coupling of the input and mixing light to the waveguide's fundamental mode of the waveguide is realized by two aspheric lenses (f = 11$\,$mm) with custom-made anti-reflection (AR) coatings for all three wavelengths. We achieve efficiencies of 90\% for the input light and 87\% for the pump laser. An internal conversion efficiency of 96.2\% from 780$\,$nm to 1522$\,$nm is achieved, mainly limited by the spatial mode overlap between pump and input light inside the waveguide. The converted light is separated by another DM from the pump laser and subsequently coupled with an efficiency of about 91\% to a single-mode telecom fiber, which has an AR-coated end facet. A band-pass filter (center wavelength: 1535$\,$nm, FWHM: 30$\,$nm, \textit{Omega optical}) removes a major part of the pump laser as well as broadband anti-Stokes Raman noise outside the pass-band. Note that the remaining parts of the filtering system are located at the polarization analysis platform for technical reasons, see Sec. \ref{sec:SpecFil}.

On this platform we achieve an external efficiency of 62.8\% (Note that the difference to 57\% is determined by the remaining spectral filters). As already mentioned in the main text, it is limited by coupling of the input light to the waveguide (90\%), internal efficiency (96.2\%), and single-mode fiber in- and out-coupling (87.8\%). Another factor which should not be underestimated, are passive losses in all optical components of 82.6\%. Although all components were optimized for the given wavelength combination, i.e. each component has at least 97\% transmission, the loss sums up to this non-negligible contribution. In particular, losses occur in the aspheric lenses for waveguide and fiber coupling, achromatic PBS/waveplate, silver mirrors in the Sagnac interferometer, band-pass filter, and dichroic mirrors.

All numbers are given for the s-polarized component; the single efficiencies of the p-polarized component may vary in the lower percent regime due to slight asymmetries in the interferometer. These asymmetries are most probably caused by slightly different cut of the end facets of the waveguide as well as chromatic dispersion in all optical components leading to a spatial displacement of the three wavelengths. The asymmetry is clearly visible in the power-dependent conversion efficiency of the two arms, which should show a perfect overlap, ideally (see Fig. 2 in the main text). Nevertheless, by careful alignment and an appropriate setting of the pump power in both arms, we achieve equal efficiencies for both polarization components.          
        
\subsection{Spectral filtering and polarization analysis} \label{sec:SpecFil}

\begin{figure}[h]
\centering
\includegraphics[width=0.85\linewidth]{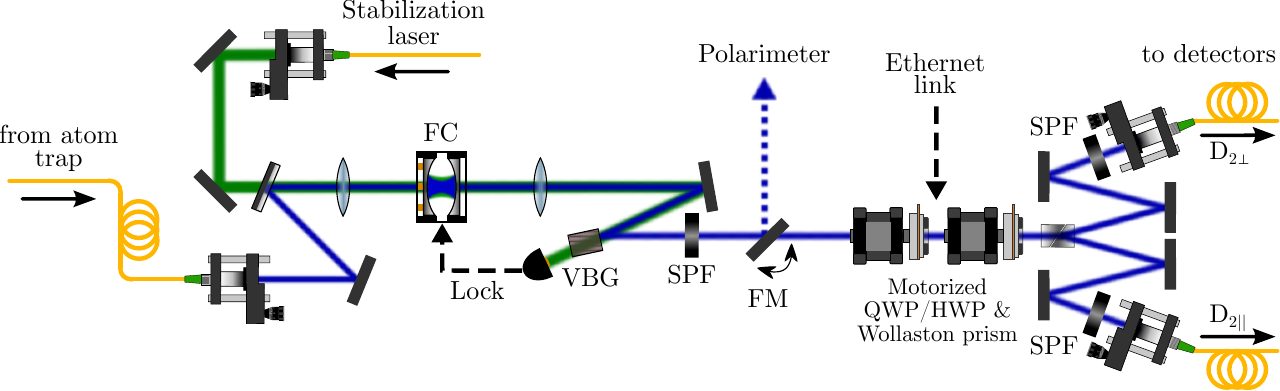}
\caption{\textbf{Setup of the spectral filtering stage and polarization analyzer.} Details on the setup can be found in the text. The following abbreviations are used: volume Bragg grating (VBG), short-pass filter (SPF), filter cavity (FC), flip mirror (FM), and quarter/half wave plate (QWP/HWP).}
\label{fig:bsm}
\end{figure}

The final platform contains the polarization analyzer as well as the remaining spectral filters. For future experiments, it can be easily modified to perform a photonic BSM with a similar scheme as in \cite{RosenfeldPRL2017}. The setup of the polarization analyzer is shown in Fig. \ref{fig:bsm}.

The spectral filters were moved to this platform mainly for two reasons. On the one hand mode-matching to the filter cavity is simplified due to a clean spatial mode coming out of the single-mode fiber between the QFC device and the polarization analyzer. On the other hand, a second identical filter cavity required for the BSM is located also on this platform. Since both cavities are in one place, technical overhead for their stabilization is reduced. The complete filter system consists of several broadband interference filters, in detail, one band-pass filter (located on the QFC platform) and two short-pass filters (1560$\,$nm cut-off), intended to remove remains of the mixing laser, Raman noise in the fibers induced by the mixing laser, and noise from unwished non-phasematched nonlinear processes (e.g. second-harmonic light of the mixing laser, etc...). A narrow-band filter formed by a volume Bragg grating (VBG) with a FWHM of 25$\,$GHz and a filter cavity (FC) aims to remove broadband ASR noise in the proximity of the target wavelength. The filter cavity is designed to have a free-spectral range of 19$\,$GHz, which matches the distance between the maximum and the first minimum of the VBG's reflection spectrum. This ensures a suitable suppression of light transmitted through neighboring longitudinal modes of the cavity. A cavity finesse of 700 results in a FWHM of 27$\,$MHz. This value reflects the tradeoff between low noise, which favors a small bandwidth due to a roughly linear scaling of the noise with the filter bandwidth, and high transmission of the emitted photons whose natural linewidth is 6.1$\,$MHz. The cavity design is similar to the transfer cavity in Sec. \ref{sec:MaLa}. Optimal mode-matching to the cavity's fundamental mode is achieved by an aspheric lens with f = 11$\,$mm to couple of of the fiber and a spherical lens with f = 250$\,$mm to focus into the cavity; this results in a transmission of 96.5\% at 1522$\,$nm at resonance.

As already mentioned in Sec. \ref{sec:MaLa}, the cavity is actively stabilized using light at 1600$\,$nm. To this end, we take advantage of the sidebands already modulated onto the mixing laser to implement a lock-in stabilization scheme. Thus, the sideband's frequency (12.5$\,$MHz) is a good tradeoff, being (I) much larger than the transfer cavity linewidth (1.7$\,$MHz) resulting in a suitable PDH signal and (II) smaller than the FC bandwidth (27$\,$MHz) to get a feasible lock-in signal. At the master laser platform (Fig. \ref{fig:lasersystem}) light for the FC stabilization first passes a bandpass filter (center wavelength: 1600$\,$nm, FWHM: 50$\,$nm) to remove ASE noise at 1522$\,$nm, which is transmitted through the FC causing additional dark counts. Next, it is coupled to a fiber-based electro-optical modulator (MPZ-LN-10, 10$\,$GHz bandwidth, \textit{iXblue}). The EOM is required as the frequencies of the mixing laser and converted light are coupled by the DFG-process, i.e. a double-resonance cannot be reached by tuning the mixing laser in total (with the exception of a fairly small number of cavity lengths). To solve this, the EOM is powered by a home-built driver, which is tunable from 150$\,$MHz to 350$\,$MHz. Now, the cavity is not stabilized to the carrier, but to one of the first- or second-order sidebands. In this situation we need to find a cavity length featuring resonances separated by a frequency within the driver's tuning range; this provides enough flexibility to find a suitable cavity length within a few FSR. In the setup the stabilization laser is coupled also to the FC's fundamental mode. The transmitted laser light is separated by the VBG from the converted light and detected with an InGaAs photodiode. Demodulation of the photodiode signal with the RF-signal, which is modulated onto the diode current, yields a dispersive error signal for a PID lock.

The transmission of the complete filter system is measured to be 90.7\% determined by transmission of the filter cavity (96.5\%), spherical lenses (98.8\%), short-pass filters (99.4\%) and the VBG's diffraction efficiency (97.4\%).

Behind the filter system, a flip mirror allows to guide classical light to a home-built polarimeter for an automatized compensation of  polarization rotations in all single-mode fibers in the setup. The design of the polarimeter is adapted from \cite{RosenfeldPRL2017}. For the experiment presented in the main text an automated polarization control was not required since the long fiber was located on a spool in an air conditioned lab.

The light passes subsequently a combination of quarter/half wave plate and a wollaston prism to perform projective measurements of the photon's polarization state in different bases. The wave plates are motorized and controlled via an Ethernet link enabling convenient remote control of the latter. Finally, the photons are coupled to single-mode fibers and guided to the SNSPDs. The projection setup has an overall transmission of 92.2\% given by the transmission through the wave plates and prism (97.8\%) and the fiber coupling (94.3\%).  

\subsection{Noise properties}

An important figure of merit for QFC devices is the amount of pump-induced noise, which limits---in addition to detector dark-counts---the accessible signal-to-noise ratio (SNR). In our device, the dominant contribution to the pump-induced noise can be attributed to anti-Stokes Raman (ASR) noise due to the comparatively small frequency difference between mixing laser and converted light of 9.5$\,$THz (ASR noise occurs in lithium niobate at least until 700$\,$cm$^{-1}$ corresponding to roughly 21$\,$THz \cite{ZaskeOE2011}).
 
\begin{figure}[h!]
\centering
\includegraphics[width=0.5\linewidth]{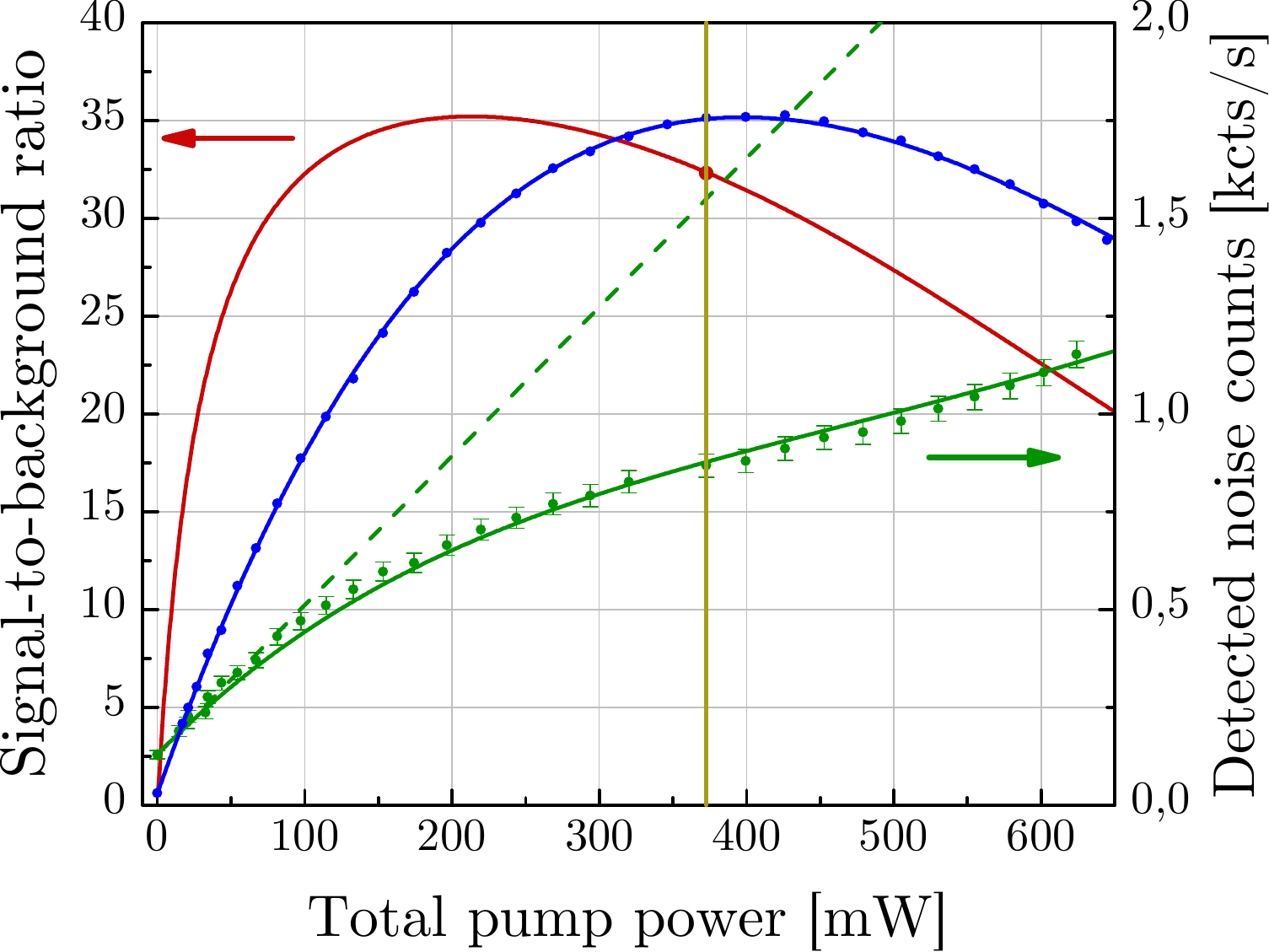}
\caption{\textbf{Noise properties of the frequency converter.} Detected noise count rates (green points), total device efficiency (blue points), and theoretically expected SNR (red line) for varying total pump power. The green and blue solid lines are fits to the data with the appropriate model (see text). The yellow vertical line indicates the operating point during the experiment.}
\label{fig:qfc_noise}
\end{figure}

The total amount of noise counts generated by the QFC device and detectors in dependence on the total pump power (i.e. the sum of both interferometer arms) is shown in Fig. \ref{fig:qfc_noise} (green data points). To this end, we measured the counts on both SNSPDs for varying pump power without 780$\,$nm input light. The detectors were operated at different bias currents than in the experiment resulting in efficiencies of 32\% and 36\%, and dark count rates of 70cps and 53cps, respectively. Following the literature \cite{ZaskeOE2011}, we would expect a linear dependence on the pump power for ASR noise. However, noise around the target wavelength can be up-converted back to 780$\,$nm subsequently after it is generated in the waveguide. Since the spectral-filter bandwidth is small compared to the acceptance bandwidth of the DFG-process (about 70$\,$GHz), phase-matching is achieved for the entire ASR noise, which results in a reduction of roughly a factor of 2 at the pump power where the internal conversion efficiency has its maximum \cite{MaringOptica2019}. To substantiate this, we start with the pump-power dependent total internal efficiency (blue data points in Fig. \ref{fig:qfc_noise}), which we fit as   
  
\begin{eqnarray}
\eta_{\text{int}}(P) = \eta_{\text{int,max}}\cdot \text{sin}^{2}\left(L\sqrt{\eta_{\text{nor}}P}\right)
\label{eq:ppdep_eff},
\end{eqnarray}

\noindent where L equals the waveguide length, $\eta_{\text{int,max}}$ equals 96.5\%, and $\eta_{\text{nor}}$ equals $1.97\frac{1}{W\cdot m^{2}}$. The detected noise count rate can be modeled as \cite{MaringOptica2019}: 

\begin{eqnarray}
N_{\text{noise}}(P) = N_{\text{dc}} + \int_{0}^{L} \alpha_{\text{ASR}}P \left(1 - \eta_{\text{int,max}}\cdot \text{sin}^{2}\left((L-x)\sqrt{\eta_{\text{nor}}P}\right)\right) \text{dx}
\label{eq:ppdep_noise},
\end{eqnarray}

\noindent taking into account a constant contribution N$_{\text{dc}}$ stemming from detector dark counts, a linear contribution $\alpha_{\text{ASR}}PL$ due to ASR noise and a nonlinear contribution due to the up-conversion of ASR noise. As shown in Fig. \ref{fig:qfc_noise} a fit to the data (solid green line) taking $N_{\text{dc}}$, $\eta_{\text{int,max}}$ and $\eta_{\text{nor}}$ as fixed parameters and the linear coefficient $\alpha_{\text{ASR}}$ as sole free parameter yields a good agreement. For comparison, the dashed green line indicates the constant and linear contribution only, i.e. disregarding the up-conversion to 780$\,$nm.

From the detected noise and device efficiency, we can model the theoretically expected behaviour of the signal-to-background ratio (red solid line) as

\begin{eqnarray}
\text{SNR} = \frac{\beta\cdot \eta_{\text{dev}}(P)}{N_{\text{noise}}(P)}
\label{eq:ppdep_SBR},
\end{eqnarray}

\noindent where $\beta$ is a scaling factor. This factor is adjusted to match the measured SNR of 32.3 (see main text) at the operating point (yellow line). As shown in Fig. \ref{fig:qfc_noise}, a slight improvement in SNR could have been achieved by reducing the pump power. However, for simplicity reasons, the converter was operated through the course of this work at maximum efficiency. 

\section{Atomic State Readout}
\label{sec:readout}

\begin{figure}
\centering
\includegraphics[width=0.35\linewidth]{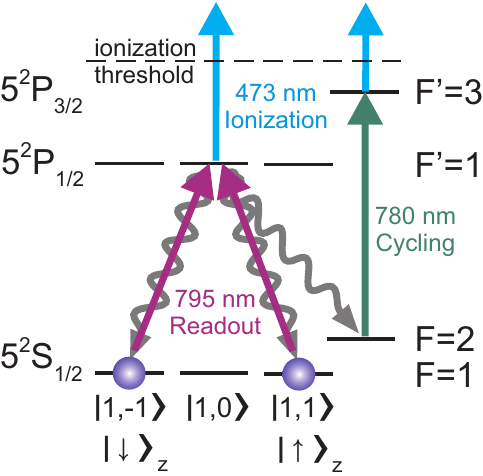}
\caption{\textbf{Atomic state measurement scheme.} A superposition of the atomic qubit state---selected by polarization of the readout light---is excited to the state $5^2 P_{1/2}$ $|F'=1, m_{F'} = 0 \rangle$ and subsequently ionized with a 473 nm laser pulse. If the atom decays to the state $5^2 S_{1/2}$ $|F=2 \rangle$ before it is ionized, a 780 nm laser pulse will transfer it to the state $5^2 P_{3/2}$ $|F'=3 \rangle$, which is ionized as well.}
\label{fig:excitation-readout}
\end{figure}

Following a successful photon detection event, the atomic spin state is analyzed using a state-selective ionization scheme \cite{krug2017thesis}, as illustrated in Fig. \ref{fig:excitation-readout}. 
The scheme starts by transferring a selected atomic qubit state superposition from the ground state $5^2 S_{1/2}$ $|F=1\rangle$ to the excited state $5^2 P_{1/2}$ $|F'=1\rangle$ with light at 795 nm (so called "readout"). Simultaneously, the excited state is ionized using ionization light at 473 nm. 
If the atom decays to the state $5^2 S_{1/2}$ $|F=2 \rangle$ before ionization, as indicated with gray arrows, a 780 nm cycling pulse will transfer it to the state $5^2 P_{3/2}$ $|F'=3 \rangle$, which is ionized as well.
The state readout is completed by fast ($<1\mu$s) \cite{ortegel2017thesis} and direct detection of the ionization fragments with channel-electron multipliers (CEM) or by fluorescence collection on a closed atomic transition to check whether the atom is still present in the trap (ionized atoms are lost).


The measurement basis is controlled by the polarization of the readout pulse, which is defined as

\begin{equation}
\chi = \cos (\alpha) \cdot V + e^{-i\phi} \sin (\alpha) \cdot H,
\end{equation}

\noindent where $V$ and $H$ denote vertical and horizontal linear light polarizations. Accordingly, two orthogonal atomic qubit state superpositions can be derived of which one is transferred to the excited state by the readout pulse (bright-state) and the other is not (dark-state), given as

\begin{equation}
\begin{split}
|\Psi\rangle_{Bright-State} & = \cos (\alpha) \frac{-1}{\sqrt{2}} (|\downarrow \rangle_z - |\uparrow \rangle_z) + \sin (\alpha) e^{i\phi} \frac{i}{\sqrt{2}} (|\downarrow \rangle_z + |\uparrow \rangle_z)  \\
|\Psi\rangle_{Dark-State} & = \sin (\alpha) \frac{1}{\sqrt{2}} (|\downarrow \rangle_z - |\uparrow \rangle_z) + \cos (\alpha) e^{i\phi} \frac{i}{\sqrt{2}} (|\downarrow \rangle_z + |\uparrow \rangle_z),
\end{split}
\end{equation}

\noindent where $|\downarrow \rangle_z$ denotes $5^2 S_{1/2}$ $|F=1, m_F=-1\rangle$ and $|\uparrow \rangle_z$ denotes $5^2 S_{1/2}$ $|F=1, m_F=+1\rangle$. Note that all states except the dark state are excited and hence ionized, e.g. population in $5^2 S_{1/2}$ $|F=1, m_F = 0 \rangle$ is always ionized. This makes the readout scheme a projection measurement onto the dark-state. 

An intuitive example of the state selectivity is the case of a $\sigma^{+}$-polarized readout pulse. In the z-basis, as shown in Fig. \ref{fig:excitation-readout}, this pulse will excite an atom in the state $|\downarrow \rangle_z$ to the state $5^2 P_{1/2}$ $|F'=1, m_{F'}=0\rangle$, however, an atom in the state $|\uparrow \rangle_z$ will not be excited because of the absence of the state $5^2 P_{1/2}$ $|F'=1, m_{F'}=2\rangle$.

\bibliographystyle{apsrev4-1}

\providecommand{\noopsort}[1]{}\providecommand{\singleletter}[1]{#1}%